\documentclass[conference,letterpaper]{IEEEtran}

\usepackage[utf8]{inputenc} 
\usepackage[T1]{fontenc}
\usepackage{url}
\usepackage{ifthen}
\usepackage{cite}
\usepackage[cmex10]{amsmath} 

\usepackage{amssymb}
\usepackage{mathrsfs}
\usepackage{bm}

\newtheorem{theorem}{Theorem}
\newtheorem{lemma}{Lemma}
\newtheorem{corollary}{Corollary}
\newtheorem{remark}{Remark}
\newtheorem{definition}{Definition}

\newtheorem{proposition}{Proposition}

\newcommand{\eqtri}{\triangleq}
\renewcommand{\IEEEQED}{\IEEEQEDopen}

\newcommand{\bx}{\bm{x}}
\newcommand{\hbx}{\hat{\bx}}
\newcommand{\cX}{\mathcal{X}}
\newcommand{\hcX}{\hat{\cX}}
\newcommand{\hX}{\hat{X}}
\newcommand{\hx}{\hat{x}}
\newcommand{\tX}{\tilde{X}}
\newcommand{\tx}{\tilde{x}}
\newcommand{\cA}{\mathcal{A}}
\newcommand{\cB}{\mathcal{B}}

\newcommand{\cG}{\mathcal{G}}
\newcommand{\cP}{\mathcal{P}}
\newcommand{\cT}{\mathcal{T}}
\newcommand{\cW}{\mathcal{W}}
\newcommand{\bcW}{\overline{\cW}}
\newcommand{\bH}{H}
\newcommand{\Ex}{\mathbb{E}}
\newcommand{\Real}{\mathbb{R}}
\newcommand{\bp}{\mathbf{p}}
\newcommand{\bq}{\mathbf{q}}

\newcommand{\rme}{\mathrm{e}}

\begin{document}
\title{Asynchronous Guessing Subject to Distortion} 

 \author{%
   \IEEEauthorblockN{Shigeaki Kuzuoka}
   \IEEEauthorblockA{Wakayama University\\
                     Email: kuzuoka@ieee.org}
 }

\maketitle

\begin{abstract}
The problem of guessing subject to distortion is considered, and the performance of randomized guessing strategies is investigated.
A one-shot achievability bound on the guessing moment (i.e., moment of the number of required queries) is given.
Applying this result to i.i.d.~sources, it is shown that randomized strategies can asymptotically attain the optimal guessing moment.
Further, a randomized guessing scheme which is feasible even when the block size is extremely large is proposed, and 
a single-letter characterization of the guessing moment achievable by the proposed scheme is obtained.
\end{abstract}


\section{Introduction}
Consider the problem of guessing the realized value $x$ of a random variable $X$ using a sequence of queries of the form ``Is $X=x$?''.
We are interested in how many queries are required until an affirmative answer is obtained; the number is called the \emph{guesswork}.
This guessing problem was introduced by Massey \cite{Massey94}, where the
expectation of guesswork was investigated.  Subsequently Arikan \cite{Arikan96}
investigated the $\rho$-th moment $G_\rho$ of guesswork, which is called the \emph{guessing moment}.
Further, Arikan and Merhav \cite{ArikanMerhav98} extended Arikan's result \cite{Arikan96} to the lossy case, 
where the value $x$ of $X$ is not necessarily identified but it is required to find a value $\hx$ satisfying $d(x,\hx)\leq\Delta$ for given distortion measure $d$ and distortion level $\Delta$.

Recently asynchronous guessing problem was introduced by Salamatian \textit{et al.}\cite{Salamatian_etal19} as an information-theoretic model for brute-force botnet attacks.
In the asynchronous setting, the guesser is restricted so that it does not know which queries were already asked.
In \cite{Salamatian_etal19}, a modified variation $V_\rho$ of the guessing moment $G_\rho$ attainable by asynchronous guessing strategies 
is investigated.
Their result implies that the optimal asynchronous guessing is given by 
\emph{randomized guessing}, where the guesser choses a query according to a certain probability distribution,
and that the penalty of lack of synchronization is asymptotically negligible.

The primary motivation of this work is to extend the study of \cite{Salamatian_etal19} to the lossy case.
Specifically, randomized guessing subject to distortion is studied.
Our model provides a simplified mathematical model for  brute-force attacks against bio-metrics authentication, where attacker's task is to find a query which is sufficiently similar to the bio-metric data stored in the system.

\subsection{Contributions}
Our first contribution is to give an achievability bound on $V_\rho$ in terms of a variation of the R\'enyi entropy (Theorem \ref{thm:main}).
In particular, when the order $\rho$ of the moment is an integer, we directly evaluate the guessing moment $G_\rho$ and give an achievability bound (Theorem \ref{thm:VandG} and its corollaries).
Our achievability result reveals that there exists a deterministic quantizer $\pi$ which does not depend on the parameter $\rho$ and 
the optimal guessing strategy is given by the tilted distribution of the quantized $\hX=\pi(X)$.

Next we apply our achievability bound to independent and identically distributed (i.i.d.) sources, and then, show that synchronization is not necessary to achieve the asymptotically
optimal guessing moment. Furthermore, for asymptotic case, we propose i.i.d.~asynchronous guessing strategies, which are simple and feasible even when the block size $n$ is extremely large.
We investigate the asymptotic performance of i.i.d.~asynchronous guessing strategies and give a single-letter characterization of the optimal guessing moment achievable by i.i.d.~strategies (Theorem \ref{thm:iid} and its corollary).

\subsection{Related Work}
The study of guessing was pioneered by Massey \cite{Massey94}.
Arikan \cite{Arikan96} demonstrated that the R\'enyi entropy \cite{Renyi61} characterizes the guessing moment (up to some factor). 
Recently, tighter bounds on the guessing moment were given by Sason and Verd\'u \cite{SasonVerdu18jun}.

The guessing problem has been studied in various contexts such as guessing allowing errors \cite{SmoothRenyi}, guessing subject to distortion
\cite{ArikanMerhav98,SaitoMatsushima_ISIT19}, investigation of large deviation perspective of guessing \cite{HanawalSundaresan11,ChristiansenDuffy13}, guesswork in multi-user systems \cite{ChristiansenDuffyPinCalmonMedard15}, 
and guesswork with distributed encoders \cite{BracherLapidothPfister19} and so on.

Applications of guessing are around the information security, e.g., cracking passwords; See the introduction of \cite{Salamatian_etal19} and Section II of \cite{MerhavCohen20} for review on guessing and security.
To understand the impact of synchronization in botnet attacks, 
Salamatian \textit{et al.}\cite{Salamatian_etal19} proposed a simplified model for distributed brute-force attacks and introduced randomized guessing.
Merhav and Cohen \cite{MerhavCohen20} studied randomized guessing under source uncertainty and proposed the universal randomized guessing strategy based on the LZ78 data compression algorithm \cite{LZ78}. 
The problem of randomized guessing under the individual-sequence approach was also investigated by Merhav \cite{Merhav20indseq}.

\subsection{Organization}
The rest of this paper is organized as follows.
In Section \ref{sec:pre}, we introduce a variation of the R\'enyi entropy and its property.
Section \ref{sec:results} describes our main results; 
one-shot results are given in 
Section \ref{sec:oneshot} and asymptotic results for i.i.d.~sources are given in 
Section \ref{sec:iid}.
All theorems are proved in Section \ref{sec:proof}.
Section \ref{sec:conclude} concludes the paper.

\section{Preliminary}\label{sec:pre}
Let $\cX$ and $\hcX$ be finite alphabets.
Let $d\colon \cX\times\hcX\to[0,\infty)$ be a distortion measure and fix the distortion level $\Delta\geq 0$.
For each $x\in\cX$, let $\cA_\Delta(x)\eqtri\{\hx\in\hcX:d(x,\hx)\leq\Delta\}$.
We assume that $\cA_\Delta(x)\neq\emptyset$ for any $x\in\cX$.

In the discussion of one-shot guessing, we will use the quantity $\bH_{\alpha}^\Delta(X)$, which was introduced in \cite{SaitoMatsushima_ISIT19}.
Let $\cW_\Delta$ be the set of conditional distributions $P_{\hX|X}$ such that $\Pr\{d(X,\hX)\leq\Delta\}=1$ (or equivalently $P_{\hX|X}(\hx|x)=0$ if $d(x,\hx)>\Delta$).
Then $\bH_{\alpha}^\Delta(X)$ is defined as follows.

\medskip
\begin{definition}
For $\alpha\in(0,1)\cup(1,\infty)$,
\begin{align}
  \bH_{\alpha}^\Delta(X)\eqtri\inf_{P_{\hX|X}\in\cW_\Delta} H_{\alpha}(\hX)
\label{eq:def_renyi}  
\end{align}
where $H_{\alpha}(\hX)$ is the R\'enyi entropy of $\hX\sim P_{\hX}$:\footnote{Throughout the paper, $\log$ denotes the natural logarithm.}
\[
H_\alpha(\hX)\eqtri\frac{1}{1-\alpha}\log\sum_{\hx\in\hcX}\left[P_{\hX}(\hx)\right]^\alpha.  
\]
\end{definition}

\medskip
As shown in Appendix \ref{sec:proof_prop:optimum_f}, the infimum in \eqref{eq:def_renyi} can be achieved by a deterministic quantizer.

\medskip
\begin{proposition}
\label{prop:optimum_f}
There exists a deterministic function $\pi\colon\cX\to\hcX$ such that $\pi(x)\in\cA_\Delta(x)$ for all $x\in\cX$ and that $\hX=\pi(X)$ satisfies, for all $\alpha\in(0,1)$,
\begin{align}
  H_{\alpha}(\hX)=\bH_{\alpha}^\Delta(X).\label{eq:cor:optimum_f}
\end{align}
\end{proposition}

\section{Main Results}\label{sec:results}
\subsection{One-Shot Bounds}\label{sec:oneshot}

Let us consider a random variable $X\sim P_X$ on $\cX$. We investigate the problem of guessing the realization value $x$ of $X$ subject to the distortion measure $d$.

An asynchronous guessing strategy is determined by a distribution $P_{\hX}$ on $\hcX$, which is independent of the realization $x$ of $X$ but may depend on $P_X$.
The guesser continues to emit i.i.d.~sequence of random variables $\hX_1,\hX_2,\dots$ according to $P_{\hX}$ as long as $d(x,\hX_i)>D$.
The number of guesses $G(x|P_{\hX})$ is given by the first index $k$ such that $d(x,\hX_k)\leq\Delta$ or equivalently $\hX_k\in\cA_\Delta(x)$.
It should be emphasized that, even when $X=x$ is fixed, the number $G(x|P_{\hX})$ of guesses is a random variable.
It is easily seen that the distribution of $G(x|P_{\hX})$ is the geometric distribution with the parameter $P_{\hX}(\cA_\Delta(x))$, i.e.,
\begin{align*}
\Pr\{G(x|P_{\hX})=k\}&=\left[1-P_{\hX}(\cA_\Delta(x))\right]^{k-1}P_{\hX}(\cA_\Delta(x)).
\end{align*}
Thus, for a given parameter $\rho>0$, the $\rho$-th moment of the number of guesses can be written as
\begin{align*}
  \Ex\left[G_\rho(X|P_{\hX})\right]=\sum_{x\in\cX}P_X(x)G_\rho(x|P_{\hX})
\end{align*}
where\footnote{Note that \(G_\rho(x|P_{\hX})\) is not a random variable although \(G(x|P_{\hX})\) is.}
\begin{align*}
G_\rho(x|P_{\hX})\eqtri\sum_{k=1}^\infty k^\rho\left[1-P_{\hX}(\cA_\Delta(x))\right]^{k-1}P_{\hX}(\cA_\Delta(x)).    
\end{align*}

While our main interest is $\Ex\left[G_\rho(X|P_{\hX})\right]$, we first investigate the quantity\footnote{The idea of considering the modification $V_\rho$ of the moment $G_\rho$ was introduced by Salamatian \textit{et al.}\cite{Salamatian_etal19}.}
\begin{align*}
  V_\rho(x|P_{\hX})\eqtri \Ex\left[
    \binom{G(x|P_{\hX})+\rho-1}{\rho}
  \right],
\end{align*}
where the expectation $\Ex$ is taken with respect to the random variable $G(x|P_{\hX})$ and $\binom{a}{b}$ is the generalized binomial coefficient defined in terms of the gamma function $\Gamma$, i.e.,
\begin{align*}
  \binom{a}{b}=\frac{\Gamma(a+1)}{\Gamma(b+1)\Gamma(a-b+1)}.
\end{align*}
The virtue of $V_\rho$ is that its value can be explicitly given as follows.

\medskip
\begin{proposition}
\label{prop:V}
For any guessing strategy $P_{\hX}$ and $\rho>0$,
\begin{align*}
  V_\rho(x|P_{\hX})=\left(\frac{1}{P_{\hX}(\cA_\Delta(x))}\right)^\rho,\quad\forall x\in\cX.
\end{align*}
\end{proposition}

\begin{corollary}
For any $P_{\hX}$ and $\rho>0$,
  \begin{align*}
    \Ex[V_\rho(X|P_{\hX})]=\sum_{x\in\cX}P_X(x)\left(\frac{1}{P_{\hX}(\cA_\Delta(x))}\right)^\rho.
    \end{align*}
\end{corollary}

\medskip
The proposition is proved in Appendix \ref{sec:proof_prop:V}.

Our one-shot achievability bound on $\Ex[V_\rho(X|P_{\hX})]$ is stated as follows.

\medskip
\begin{theorem}
\label{thm:main}
There exists a deterministic function $\pi\colon\cX\to\hcX$ such that, for all $\rho>0$,
the tilted distribution
\begin{align}
  P_{\hX_\rho^*}(\hx)\eqtri\frac{P_{\hX}(\hx)^{\frac{1}{1+\rho}}}{\sum_{\hx'}P_{\hX}(\hx')^{\frac{1}{1+\rho}}}
\label{eq:tited_distrib}
\end{align}
of  the distribution $P_{\hX}$ of $\hX=\pi(X)$ satisfies
\begin{align*}
    \log \Ex[V_\rho(X|P_{\hX_\rho^*})]\leq \rho\bH_{\frac{1}{1+\rho}}^\Delta(X).
  \end{align*}
\end{theorem}

\medskip
The theorem is proved in Section \ref{sec:proof_main}.

Now we investigate our main interest, i.e., the $\rho$-th moment $\Ex[G_\rho(X|P_{\hX})]$ of the number of guesses.
In particular, we consider the case where $\rho=1,2,\dots$ is a positive integer.%
\footnote{For non-integer \(\rho>0\), we may numerically evaluate the \(\rho\)-th moment by using the technique recently developed in \cite{MerhavSason20}.}
In this case, we can directly evaluate $G_\rho(x|P_{\hX})$ by using the moment generating function $M(t)=p_x\rme^t/(1-(1-p_x)\rme^t)$ of the geometric distribution with the parameter $p_x\eqtri P_{\hX}(\cA_\Delta(x))$; 
e.g., the first four moments are
\begin{align*}
  G_1(x|P_{\hX})&=1/p_x,\\
  G_2(x|P_{\hX})&=(2-p_x)/p_x^2,\\
  G_3(x|P_{\hX})&=(p_x^2-6p_x+6)/p_x^3,\\
  G_4(x|P_{\hX})&=(-p_x^3+14p_x^2-36p_x+24)/p_x^4.
\end{align*}
Further, we have upper and lower bounds on $G_\rho(x|P_{\hX})$ as follows.

\medskip
\begin{theorem}
\label{thm:VandG}
For any guessing strategy $P_{\hX}$, any $x\in\cX$, and any positive integer $\rho$, 
\begin{align*}
  V_\rho(x|P_{\hX})\leq
  G_\rho(x|P_{\hX})
  \leq
  (\rho!)V_\rho(x|P_{\hX}),
\end{align*}    
or equivalently
\begin{align*}
\left(\frac{1}{P_{\hX}(\cA_\Delta(x))}\right)^\rho
\leq G_\rho(x|P_{\hX})\leq(\rho!)\left(\frac{1}{P_{\hX}(\cA_\Delta(x))}\right)^\rho.
\end{align*}    
\end{theorem}

\begin{corollary}\label{cor:AandG}
For any $P_{\hX}$ and positive integer $\rho$,
  \begin{align*}
    \Ex[V_\rho(X|P_{\hX})]\leq
    \Ex[G_\rho(X|P_{\hX})]
    \leq
    (\rho!)\Ex[V_\rho(X|P_{\hX})]
  \end{align*}    
  and
  \begin{align*}
  \lefteqn{\sum_{x\in\cX}P_X(x)\left(\frac{1}{P_{\hX}(\cA_\Delta(x))}\right)^\rho}\\
  &\leq\Ex[G_\rho(X|P_{\hX})]\\
  &\leq(\rho!)\sum_{x\in\cX}P_X(x)\left(\frac{1}{P_{\hX}(\cA_\Delta(x))}\right)^\rho.
  \end{align*}  
\end{corollary}
  
\medskip
The theorem is proved in Section \ref{sec:proof_VandG}.

From Theorems \ref{thm:main} and \ref{thm:VandG}, we can obtain a one-shot achievability result in terms of $\Ex[G(X|P_{\hX})^\rho]$ as follows.
\medskip

\begin{corollary}\label{cor:one-shot-achievability}
There exists a deterministic function $\pi\colon\cX\to\hcX$ such that, for any positive integer $\rho$,
the tilted distribution $P_{\hX_\rho^*}$ defined as \eqref{eq:tited_distrib} satisfies
\begin{align*}
  \log \Ex[G_\rho(X|P_{\hX_\rho^*})]\leq \rho\bH_{\frac{1}{1+\rho}}^\Delta(X)+\log(\rho!).
\end{align*}
\end{corollary}

\medskip
Let us compare our result with that of synchronous case.
A synchronous guessing strategy is determined by a bijection
$\cG\colon\hcX\to\{1,2,\dots,|\hcX|\}$, and the number of guesses when $X=x$ is given by
\[
  G^{\textsf{sync}}(x|\cG)\eqtri\min_{\hx\in\cA_\Delta(x)}\cG(\hx).
\]
According to \cite{SaitoMatsushima_ISIT19}, the optimal $\rho$-th moment achievable by synchronous strategies 
satisfies
\begin{align}
\lefteqn{\rho\bH_{\frac{1}{1+\rho}}^\Delta(X)-\rho\log\log(1+\min\{|\cX|,|\hcX|\})}\nonumber\\
&\leq \log \min_{\cG}\Ex[G^{\textsf{sync}}(X|\cG)^\rho]\nonumber\\
&\leq \rho\bH_{\frac{1}{1+\rho}}^\Delta(X).\label{eq:synchronous}
\end{align}
Comparing \eqref{eq:synchronous} with Corollary \ref{cor:one-shot-achievability}, we can see that the penalty of lack of synchronization is upper bounded by
\begin{align*}
 \log(\rho!)+\rho\log\log(1+\min\{|\cX|,|\hcX|\}).
\end{align*}

\subsection{Asymptotics for Stationary Memoryless Sources}\label{sec:iid}
In this subsection, 
we apply our one-shot results to i.i.d.~sources and investigate the asymptotic behavior of the $\rho$-th moment of the number of guesses.\footnote{To simplify the argument, we assume that $\rho$ is a positive integer. However, it is not hard to show that our argument is valid for non-integer $\rho>0$; See Appendix \ref{sec:non-integer}.}

Let $\cX^n$ (resp.~$\hcX^n$) is the $n$-fold Cartesian product of $\cX$ (resp.~$\hcX$).
The distortion between $\bx\in\cX^n$ and $\hbx\in\hcX^n$ per symbol is defined by
$d_n(\bx,\hbx)=(1/n)\sum_{i=1}^nd(x_i,\hx_i)$.
We investigate the problem of guessing the realization value $\bx$ of $X^n=(X_1,X_2,\dots,X_n)$ subject to the distortion measure $d_n$, where $X_1,\dots,X_n$ are independently generated according to an identical distribution $P_X$ on $\cX$.

As in the one-shot case, an asynchronous guessing strategy is determined by a distribution $P_{\hX^n}$ on $\hcX^n$.
As a direct consequence of Corollary \ref{cor:one-shot-achievability}, there exists 
a deterministic function $\pi_n\colon\cX^n\to\hcX^n$ such that the strategy $P_{\hX_\rho^{*n}}$ induced by $\pi_n$ satisfies
\begin{align*}
\frac{1}{n}\log\Ex[G_\rho(X^n|P_{\hX_\rho^{*n}})] 
\leq\frac{1}{n}\rho\bH_{\frac{1}{1+\rho}}^\Delta(X^n)+\zeta_n
\end{align*}
where $\zeta_n\eqtri (\rho!)/n\to 0$ as $n\to\infty$.
This fact indicates that synchronization is not necessary to achieve the asymptotically optimal guessing moment. 

However, the strategy $P_{\hX_\rho^{*n}}$ may be not feasible when $n$ is large.
In particular, it may not be easy to find and implement the function $\pi_n$. 
Hence, we restrict the class of guessing strategies.

\medskip
\begin{definition}
An asynchronous guessing strategy $P_{\hX^n}$ is said to be an \emph{i.i.d.~asynchronous guessing strategy} if there exists a distribution $Q_{\hX}$ on $\hcX$ satisfying 
\begin{align*}
  P_{\hX^n}(\hbx)=Q_{\hX}^n(\bx)\eqtri\prod_{i=1}^nQ_{\hX}(\hx_i),\quad\forall\hbx\in\hcX^n.
\end{align*}
\end{definition}

\medskip
In the following, we investigate the optimal guessing moment 
asymptotically achievable by i.i.d.~asynchronous guessing strategies. To state our result, we introduce some notation. 
We use the following standard information-theoretic quantities \cite{CsiszarKorner2nd}.
For a distribution $P$ and a conditional distribution $V$, let 
$H(P)$ be the entropy of $P$, 
$H(V|P)=\sum_{x}P(x)H(V(\cdot|x))$ be the conditional entropy, and 
$I(P,V)=H(PV)-H(V|P)$ be the mutual information, where 
$PV$ is the distribution such that $PV(\hx)=\sum_xP(x)V(\hx|x)$.
For two distributions $P$ and $Q$, let $D(P\Vert Q)$ be the divergence between $P$ and $Q$.
Let $\bcW_\Delta(Q_X)$ be the set of conditional distributions satisfying $\sum_{x,\hx}Q_X(x)V(\hx|x)d(x,\hx)\leq\Delta$.

\medskip
\begin{definition}
For distributions $Q_X$ on $\cX$ and $Q_{\hX}$ on $\hcX$, 
\begin{align*}
R(Q_X,Q_{\hX}|\Delta)\eqtri \min_{V\in\bcW_\Delta(Q_X)}[I(Q_X,V)+D(Q_XV\Vert Q_{\hX})]
\end{align*}
where $Q_XV$ is the distribution such that $Q_XV(\hx)=\sum_xQ_X(x)V(\hx|x)$.
\end{definition}

\medskip
The next theorem, which is proved in Section \ref{sec:proof_iid}, is our main result of this subsection.

\begin{theorem}\label{thm:iid}
For any i.i.d.~asynchronous guessing strategy $Q_{\hX}$ and positive integer $\rho$,
\begin{align*}
\lefteqn{\lim_{n\to\infty}\frac{1}{n}\log\Ex[G_\rho(X^n|Q_{\hX}^n)] }\\
&= \max_{Q_X}\left[\rho R(Q_X,Q_{\hX}|\Delta)-D(Q_X\Vert P_X)\right]
\end{align*}
where the maximum is taken over all distributions $Q_X$ on $\cX$.
\end{theorem}

\medskip
As a consequence of the theorem, we obtain the following result on 
the optimal $\rho$-th moment achievable by i.i.d.~asynchronous guessing strategies.

\medskip
\begin{corollary}
The exponent of the optimal $\rho$-th moment achievable by i.i.d.~asynchronous guessing strategies is 
\begin{align*}
  E_\rho^{\textsf{i.i.d}}(P_X|\Delta)\eqtri\min_{Q_{\hX}}\max_{Q_X}\left[\rho R(Q_X,Q_{\hX})-D(Q_X\Vert P_X)\right]
\end{align*}
where $\min$ (resp.~$\max$) is taken over all distributions $Q_{\hX}$ on $\hcX$ (resp.~$Q_X$ on $\cX$).
\end{corollary}

\smallskip

\begin{remark}
The corollary guarantees that we can find the optimal i.i.d.~strategy by solving the minimization in the definition of $E_\rho^{\textsf{i.i.d}}(P_X|\Delta)$, which does not depend on $n$.
So, our strategy is feasible even when $n$ is extremely large.
\end{remark}

\smallskip

\begin{remark}
  It should be emphasized that Theorem \ref{thm:iid} holds for any strategy $Q_{\hX}$.
  In other words, $Q_{\hX}$ is not necessarily depend on $P_X$.
  Hence, it can be easily applied to guessing under source uncertainty. 
  Assume that the guesser does not know the source distribution $P_X$ but it knows the fact that $P_X\in\cP$ for a subset $\cP$ of distributions.
  Theorem \ref{thm:iid} shows that, under this setting, the exponent of the optimal $\rho$-th guessing moment asymptotically achievable by i.i.d.~strategies is
  \begin{align*}
    \min_{Q_{\hX}}\max_{P_X\in\cP}\max_{Q_X}\left[\rho R(Q_X,Q_{\hX})-D(Q_X\Vert P_X)\right].
  \end{align*}
\end{remark}

\medskip
Lastly, we investigate the penalty of restricting strategies to be i.i.d.

Let us define
\begin{align*}
  E_\rho(P_X|\Delta)\eqtri\max_{Q_X}\left\{\rho R(Q_X|\Delta)-D(Q_X\Vert P_X)\right\}
\end{align*}
where the maximum is taken over all distributions $Q_X$ on $\cX$
and $R(Q_X|\Delta)$ is the \emph{rate-distortion function}; i.e.,
\begin{align*}
  R(Q_X|\Delta)\eqtri \min_{V\in\bcW_\Delta(Q_X)} I(Q_X,V).
\end{align*}
It is known that $E_\rho(P_X|\Delta)$ is the exponent of the optimal $\rho$-th guessing moment asymptotically achievable by synchronous strategies
\cite{ArikanMerhav98}; i.e.,
\begin{align*}
E_\rho(P_X|\Delta)&=\lim_{n\to\infty}\frac{1}{n}\log \min_{\cG_n\text{ on }\hcX^n}\Ex[G^{\textsf{sync}}(X^n|\cG_n)^\rho].
\end{align*}
Further, results of \cite{SaitoMatsushima_ISIT19} implies that
\begin{align*}
  E_\rho(P_X|\Delta)&=\lim_{n\to\infty}\frac{\rho}{n}H_{\frac{1}{1+\rho}}^\Delta(X^n).
\end{align*}

On the other hand, since
\begin{align*}
  \min_{Q_{\hX}}R(Q_X,Q_{\hX}|\Delta)=R(Q_X|\Delta),
\end{align*}
we have
\begin{align*}
  E_\rho^{\textsf{i.i.d}}(P_X|\Delta)
&=\min_{Q_{\hX}}\max_{Q_X}\left[\rho R(Q_X,Q_{\hX}|\Delta)-D(Q_X\Vert P_X)\right]\\
&\geq \max_{Q_X}\min_{Q_{\hX}}\left[\rho R(Q_X,Q_{\hX}|\Delta)-D(Q_X\Vert P_X)\right]\\
&= E_\rho(P_X|\Delta).
\end{align*}
From this, we can see the suboptimality of i.i.d.~strategies and evaluate the penalty as
\begin{align*}
\lefteqn{\min_{Q_{\hX}}\max_{Q_X}\left[\rho R(Q_X,Q_{\hX}|\Delta)-D(Q_X\Vert P_X)\right]}\\
&-\max_{Q_X}\min_{Q_{\hX}}\left[\rho R(Q_X,Q_{\hX}|\Delta)-D(Q_X\Vert P_X)\right].
\end{align*}

\section{Proofs}\label{sec:proof}

\subsection{Proof of Theorem \ref{thm:main}}\label{sec:proof_main}
Let $\pi$ be the function given in Proposition \ref{prop:optimum_f} and let $\hX=\pi(X)$.
Since $\pi(x)\in\cA_\Delta(x)$, we have $P_{\hX_\rho^*}(\cA_\Delta(x))\geq P_{\hX_\rho^*}(\pi(x))$.
Hence, letting $\pi^{-1}(\hx)\eqtri\{x\in\cX: \pi(x)=\hx\}$,
we have
\begin{align*}
  \Ex[V_\rho(X|P_{\hX_\rho^*})]
&=\sum_{x\in\cX}P_X(x)\left(\frac{1}{P_{\hX_\rho^*}(\cA_\Delta(x))}\right)^\rho\\
&\leq  \sum_{x\in\cX}P_X(x)\left(\frac{1}{P_{\hX_\rho^*}(\pi(x))}\right)^\rho\\
&=\sum_{\hx\in\hcX}  \sum_{x\in \pi^{-1}(\hx)}P_X(x)\left(\frac{1}{P_{\hX_\rho^*}(\hx)}\right)^\rho\\
&=\sum_{\hx\in\hcX}  P_{\hX}(\hx)\left(\frac{1}{P_{\hX_\rho^*}(\hx)}\right)^\rho\\
&= \exp\left[\rho H_{\frac{1}{1+\rho}}(\hX)\right]\\
&= \exp\left[\rho\bH_{\frac{1}{1+\rho}}^\Delta(X)\right]
\end{align*}
where the last equality follows from \eqref{eq:cor:optimum_f}.
\hfill\IEEEQED

\subsection{Proof of Theorem \ref{thm:VandG}}\label{sec:proof_VandG}

  Since $\rho>1$, using Jensen's inequality, we have
  \begin{align}
    G_\rho(x|P_{\hX})
  &=\sum_{k=1}^\infty \Pr\{G(x|P_{\hX})=k\}k^\rho\nonumber\\
  &\geq \left\{\sum_{k=1}^\infty \Pr\{G(x|P_{\hX})=k\}k\right\}^\rho\nonumber\\
  &=\{1/P_{\hX}(\cA_\Delta(x))\}^{\rho}\label{eq:tmp1_proof_thm:VandG}
  \end{align}
  where the last equality follows from the fact that 
  the distribution of $G(x|P_{\hX})$ is the geometric distribution with the parameter $P_{\hX}(\cA_\Delta(x))$.
  
  On the other hand, since $G(x|P_{\hX})$ is an integer-valued random variable and $\rho$ is an integer, 
  \begin{align*}
    \frac{G(x|P_{\hX})^\rho}{\rho!}
  &\leq \frac{1}{\rho!}G(x|P_{\hX})\times[G(x|P_{\hX})+1]\\
  &\quad\times[G(x|P_{\hX})+2]\times\dots\times[G(x|P_{\hX})+\rho-1]\\
  &=\binom{G(x|P_{\hX})+\rho-1}{\rho}.
  \end{align*}
  Taking the expectation with respect to $G(x|P_{\hX})$ and multiply both sides by $\rho!$, we have
  \begin{align}
    G_\rho(x|P_{\hX})\leq (\rho!) V_\rho(x|P_{\hX}).\label{eq:tmp2_proof_thm:VandG}
  \end{align}

  Combining \eqref{eq:tmp1_proof_thm:VandG} and \eqref{eq:tmp2_proof_thm:VandG}
with Proposition \ref{prop:V}, we have the theorem.
\hfill\IEEEQED 

\subsection{Proof of Theorem \ref{thm:iid}}\label{sec:proof_iid}
Before giving the proof, we introduce some notation.

For two positive sequences $a_n$ and $b_n$, we will write 
$a_n\doteq b_n$ if $\lim_{n\to\infty}(1/n)\log(a_n/b_n)=0$.
Similarly, when $a_n$ and $b_n$ depend on a sequence $\bx$, the notation $a_n(\bx)\doteq b_n(\bx)$ means that
\begin{align*}
  \lim_{n\to\infty}\max_{\bx\in\cX^n}\left|\frac{1}{n}\log\frac{a_n(\bx)}{b_n(\bx)}\right|=0.
\end{align*}

In our proof, we use the \emph{method of types} \cite{CsiszarKorner2nd}.
For $\bx\in\cX^n$, the \emph{type} $P_{\bx}$ is the empirical distribution of $\bx=(x_1,\dots,x_n)$; i.e.,
$P_{\bx}(a)=(1/n)|\{1\leq i\leq n: x_i=a\}|$ for all $a\in\cX$.
Let $\cP_n$ be the possible types of length $n$ sequences. For $Q\in\cP_n$, $\cT_Q$ is the set of sequences $\bx$ such that $P_{\bx}=Q$.
For a conditional distribution $V\colon\cX\to\hcX$ and $\bx\in\cX^n$,
$\cT_V(\bx)$ denotes the set of sequences $\hbx=(\hx_1,\dots,\hx_n)\in\hcX^n$ satisfying $V(b|a)=|\{j:(x_j,\hx_j)=(a,b)\}|/|\{i:x_i=a\}|$ for all $(a,b)\in\cX\times\hcX$.

\begin{IEEEproof}[Proof of Theorem \ref{thm:iid}]
From Corollary \ref{cor:AandG}, we have
\begin{align}
  \Ex[G_\rho(X^n|Q_{\hX}^n)]\doteq 
  \sum_{\bx\in\cX^n}P_{X^n}(\bx)[Q_{\hX}^n(\cA_\Delta(\bx))]^{-\rho}\label{eq:tmp1_proof_iid}
\end{align}
where $\cA_\Delta(\bx)\eqtri\{\hbx\in\hcX^n:d_n(\bx,\hbx)\leq\Delta\}$.
So, we evaluate the exponent of the right-hand side of \eqref{eq:tmp1_proof_iid}. 

For any $\hbx$ and $V$, we have
\begin{align*}
    Q_{\hX}^n(\hbx)&=\exp\left\{-n[H(P_{\hbx})+D(P_{\hbx}\Vert Q_{\hX})]\right\},\\
    |\cT_V(\bx)|&\doteq\exp\{nH(V|P_{\bx})\}
\end{align*}
and thus,  
\begin{align*}
    Q_{\hX}^n(\cT_V(\bx))\doteq\exp\left\{-n[I(P_{\bx},V)+D(P_{\bx}V\Vert Q_{\hX})]\right\}
\end{align*}
where $P_{\bx}V$ is the distribution on $\hcX$ such that $P_{\bx}V(\hx)=\sum_xP_{\bx}(x)V(\hx|x)$.

Further, $\cA_\Delta(\bx)$ can be written as
\begin{align*}
    \cA_\Delta(\bx)=\bigcup_{\substack{V:\\ \sum_{x,\hx}P_{\bx}(x)V(\hx|x)d(x,\hx)\leq\Delta}}\cT_V(\bx).
\end{align*}
Hence, we have
\begin{align}
    Q_{\hX}^n(\cA_\Delta(\bx))&\doteq\exp\left\{-n\min_{V}[I(P_{\bx},V)+D(P_{\bx}V\Vert Q_{\hX})]\right\}\nonumber\\
  &=\exp\left\{-nR(P_{\bx},Q_{\hX}|\Delta)\right\}.\label{eq:tmp_proof_iid}
\end{align}

On the other hand, we have 
\[P_X(\cT_{Q_X})\doteq\exp\{-nD(Q_X\Vert P_X)\}
\]
for all $Q_X\in\cP_n$. Combining this with \eqref{eq:tmp_proof_iid}, we have
\begin{align*}
    \lefteqn{\sum_{\bx\in\cX^n}P_{X^n}(\bx)[Q_{\hX}^n(\cA_\Delta(\bx))]^{-\rho}}\\
    &=\sum_{Q_X\in\cP_n}\sum_{\bx\in\cT_{Q_X}}P_{X^n}(\bx)[Q_{\hX}^n(\cA_\Delta(\bx))]^{-\rho}\\
    &\doteq\sum_{Q_X\in\cP_n}\exp\left\{n\left[\rho R(Q_X,Q_{\hX}|\Delta)-D(Q_X\Vert P_X)\right]\right\}\\
    &\doteq\exp\left\{n\max_{Q_X}\left[\rho R(Q_X,Q_{\hX}|\Delta)-D(Q_X\Vert P_X)\right]\right\}.
\end{align*}
\end{IEEEproof}

\section{Concluding Remarks}\label{sec:conclude}
In this paper, randomized strategies for guessing subject to distortion was studied.
A one-shot achievability bound on the guessing moment was given.
Further, feasible i.i.d.~asynchronous guessing scheme was proposed, and its asymptotic performance was investigated.

Lastly, we give some comments regarding generalizations of our results.
\begin{itemize}
  \item It is not hard to extend the result to the case where side-information is available at the guesser.
  \item Our result shows that the behavior of 
  \[
  -\frac{1}{n}\log Q_{\hX}^n(\cA_\Delta(\bx))\simeq R(P_{\bx},Q_{\hX}|\Delta)
  \]
  determines the guessing moment (See \eqref{eq:tmp_proof_iid} and \eqref{eq:tmp1_proof_iid} in the proof of Theorem \ref{thm:iid}).
  Since the behavior of $(-1/n)\log Q_{\hX}^n(\cA_\Delta(\bx))$ for sources with memory is well studied in the context of the rate-distortion theory (see \cite{DemboKontoyiannis02} and references there in), we can apply those results.
  For example, our argument can also be applied to stationary ergodic sources by using Theorem 3 of \cite{YangZhang99May}.
\end{itemize}

\appendices
\section{Proof of Proposition \ref{prop:optimum_f}}\label{sec:proof_prop:optimum_f}
First we introduce the concept of majorization and Schur concavity, which play important role in the proof.

Let $\Real_+^m$ be the set of vectors with $m$ nonnegative components.
Given $\bp=(p_1,p_2,\dots,p_m)\in\Real_+^m$, denote by $p_{[1]}\geq p_{[2]}\geq \dots\geq p_{[m]}$ the permutation of the components of $\bp$ in the nonincreasing order.

\begin{definition}
We say that $\bq\in\Real_+^m$ majorizes $\bp\in\Real_+^m$ (and write $\bp\prec\bq$) if 
\begin{align*}
  \sum_{i=1}^jp_{[i]}\leq \sum_{i=1}^jq_{[i]},\quad \forall j=1,2,\dots,m-1
\end{align*}
and
\begin{align}
  \sum_{i=1}^mp_{[i]}\leq \sum_{i=1}^mq_{[i]}.
\end{align}
\end{definition}

\begin{definition}
A real valued function $h$ on $\Real_+^m$ is said to be \emph{Schur concave} if $h(\bp)\geq h(\bq)$ for any $\bp,\bq\in\Real_+^m$ satisfying $\bp\prec\bq$.
\end{definition}

\medskip
It is well known that $\phi(\bp)=\sum_{i=1}^m (h_i)^\alpha$ for $\alpha\in(0,1)$ is Schur concave; See, e.g.~\cite{Marshall1979Inequalities}.
Thus, we can easily see that the R\'enyi entropy of order $\alpha\in(0,1)$ is also Schur concave.
Hence, to prove Proposition \ref{prop:optimum_f}, it is sufficient to prove the following lemma.
Although the same argument is given in the last page of \cite{SaitoYagiMatsushima_ISIT17}, we give a proof for the completeness.

\begin{lemma}
  There exists a deterministic function $\pi\colon\cX\to\hcX$ such that 
  (i) $d(x,\pi(x))\leq\Delta(x)$ for all $x\in\cX$ and (ii) the distribution $P_{\hX}$ of $\hX=\pi(X)$
  majorizes any $P_{\tX}$ induced by $P_{\tX|X}\in\cW_\Delta$.
\end{lemma}

\begin{IEEEproof}
  Let $m\eqtri|\cX|$ in this proof. For each $\hx\in\cX$, let
  \begin{align*}
    \cB_\Delta(\hx)\eqtri\{x\in\cX:d(x,\hx)\leq\Delta\}.
  \end{align*}
  We define the order $\hx_1,\hx_2,\dots,\hx_m$ of symbols in $\hcX$ as follows.
  Let $\hx_1$ be a symbol satisfying
  \begin{align*}
    \Pr\{X\in\cB_\Delta(\hx_1)\}=\max_{\hx\in\hcX}\Pr\{X\in\cB_\Delta(\hx)\}.
  \end{align*}
  Then, for $i=2,3,\dots,m$, let $\hx_i\in\hcX\setminus\{\hx_1,\dots,\hx_{i-1}\}$ be a symbol such that
  \begin{align*}
  \lefteqn{\Pr\left\{X\in\cB_\Delta(\hx_i)\setminus\bigcup_{j=1}^{i-1}\cB_\Delta(\hx_j)\right\}}\\
  &=\max_{\hx\in \hcX} \Pr\left\{X\in\cB_\Delta(\hx)\setminus\bigcup_{j=1}^{i-1}\cB_\Delta(\hx_j)\right\}.
  \end{align*}
  Let $\cX_i\eqtri \cB_\Delta(\hx_i)\setminus\bigcup_{j=1}^{i-1}\cB_\Delta(\hx_j)$ ($i=1,\dots,m$).
  Then $\cX_1,\dots,\cX_m$ give a partition of $\cX$, and thus, we can define $\pi\colon\cX\to\hcX$ so that $\pi(x)=\hx_i$ if $x\in\cX_i$.
  It is apparent that $d(x,\pi(x))\leq\Delta$ for all $x\in\cX$.
  Further, from the construction, the distribution $P_{\hX}$ of $\hX=\pi(X)$ satisfies
  \begin{align*}
    P_{\hX}(\hx_1)\geq P_{\hX}(\hx_2)\geq\dots\geq P_{\hX}(\hx_{m}).
  \end{align*}
  
  We will prove that $\pi$ satisfies (ii) by contradiction.
  Assume that there exists $P_{\tX|X}\in\cW_\Delta$ such that $P_{\tX}$ induced by $P_{\tX|X}$ is \emph{not} majorized by $P_{\hX}$.
  Let us define another order $\tx_1,\tx_2,\dots,\tx_m$ in $\hcX$ so that
  \begin{align*}
    P_{\tX}(\tx_1)\geq P_{\tX}(\tx_2)\geq\dots\geq P_{\tX}(\tx_{m}).
  \end{align*}
  Since $P_{\tX}\not\prec P_{\hX}$, there exists $k$ such that
  \begin{align*}
    \sum_{i=1}^k P_{\tX}(\tx_i)> \sum_{i=1}^k P_{\hX}(\hx_i).
  \end{align*}
  Hence, we have
  \begin{align*}
  \lefteqn{\Pr\{d(X,\tX)>\Delta\}}\\
  &\geq \sum_{x\in\cX}\sum_{i=1}^kP_X(x)P_{\tX|X}(\tx_i|x)\bm{1}[d(x,\tx_i)>\Delta]\\
  &=\sum_{i=1}^kP_{\tX}(\tx_i)\\
  &\quad -\sum_{x\in\cX}P_X(x)\sum_{i=1}^kP_{\tX|X}(\tx_i|x)\bm{1}[x\in\cB_\Delta(\tx_i)]\\
  &\stackrel{\text{(a)}}{\geq}\sum_{i=1}^kP_{\tX}(\tx_i)-\Pr\left\{X\in\bigcup_{i=1}^k\cB_\Delta(\tx_i)\right\}\\
  &\stackrel{\text{(b)}}{\geq} \sum_{i=1}^kP_{\tX}(\tx_i)-\Pr\left\{X\in\bigcup_{i=1}^k\cB_\Delta(\hx_i)\right\}\\
  &= \sum_{i=1}^kP_{\tX}(\tx_i)-\sum_{i=1}^k\Pr\left\{X\in\cX_i\right\}\\
  &= \sum_{i=1}^kP_{\tX}(\tx_i)-\sum_{i=1}^k P_{\hX}(\hx_i)\\
  &>0,
  \end{align*}
  where $\bm{1}$ denotes the indicator function, 
  (a) follows from $\sum_{i=1}^kP_{\tX|X}(\tx_i|x)\bm{1}[x\in\cB_\Delta(\tx_i)]\leq \bm{1}[x\in\bigcup_{i=1}^k\cB_\Delta(\tx_i)]$ for all $x\in\cX$,
  and (b) follows from the definition of the order $\hx_1,\hx_2,\dots,\hx_m$.
  This contradicts the fact $P_{\tX|X}\in\cW_\Delta$.
\end{IEEEproof}

\section{Proof of Proposition \ref{prop:V}}\label{sec:proof_prop:V}
The proposition can be proved in the same manner as \cite[Lemma 2]{Salamatian_etal19}. We give a proof for the completeness.

\begin{IEEEproof}[Proof of Proposition \ref{prop:V}]
Letting $p_x\eqtri P_{\hX}(\cA_\Delta(x))$, we have
\begin{align*}
  V_\rho(x|P_{\hX})
  &=\sum_{m=1}^\infty\binom{m+\rho-1}{\rho}\Pr\{G(x|P_{\hX})=m\}\\
  &=p_x \sum_{m=1}^\infty\binom{m+\rho-1}{\rho}(1-p_x)^{m-1}\\
  &\stackrel{\text{(a)}}{=}p_x \sum_{m=1}^\infty\binom{-\rho-1}{m-1}[-(1-p_x)]^{m-1}\\
  &=p_x \sum_{k=0}^\infty\binom{-\rho-1}{k}[-(1-p_x)]^{k}\\
  &\stackrel{\text{(b)}}{=}p_x [1-(1-p_x)]^{-\rho-1}\\
  &=(p_x)^{-\rho}
\end{align*}
where (a) follows from
the relationship
\begin{align*}
  \binom{m+\rho-1}{\rho}=(-1)^{m-1}\binom{-\rho-1}{m-1},
\end{align*}
which is proved in the proof of Lemma 2 in \cite{Salamatian_etal19}, and (b) follows from the binominal formula.
\end{IEEEproof}

\section{Asymptotics for non-integer $\rho$}\label{sec:non-integer}
We show that Theorem \ref{thm:iid} also holds for non-integer $\rho>0$.
From \eqref{eq:tmp_proof_iid}, for all $\bx\in\cX^n$ and $Q_{\hX}$,
\begin{align*}
  q_{\bx}\eqtri Q_{\hX}^n(\cA_\Delta(\bx))\doteq\exp\left\{-nR(P_{\bx},Q_{\hX}|\Delta)\right\}.
\end{align*}
Assume that $R(P_{\bx},Q_{\hX}|\Delta)>0$.
Then, since $q_{\bx}<1/2$ for large $n$,
(20) of \cite{Merhav20indseq} gives
\begin{align*}
  G_\rho(\bx|Q_{\hX}^n)&=\sum_{k=1}^\infty k^\rho (1-q_{\bx})^{k-1}q_{\bx}\\
  &\geq \left(\frac{1-q_{\bx}}{q_{\bx}}\right)^\rho\exp\left\{-\frac{1}{1-q_{\bx}}\right\}\\
  &\geq \frac{2^{-\rho}}{\rme^2}q_{\bx}^{-\rho}.
\end{align*}
Thus, we have
\begin{align}
  \liminf_{n\to\infty}\frac{1}{n}\log
\left|\frac{G_\rho(\bx|Q_{\hX}^n)}{\exp\{\rho n R(P_{\bx},Q_{\hX}|\Delta)\}}\right|
\geq 0.
\label{eq1:non-integer}
\end{align}
On the other hand, by using Lemma 1 of \cite{MerhavCohen20} with $a=R(P_{\bx},Q_{\hX}|\Delta)$, we can show that
\begin{align}
  \limsup_{n\to\infty}\frac{1}{n}\log
\left|\frac{G_\rho(\bx|Q_{\hX}^n)}{\exp\{\rho n R(P_{\bx},Q_{\hX}|\Delta)\}}\right|
\leq 0. 
\label{eq2:non-integer}
\end{align}
Combining \eqref{eq1:non-integer} and \eqref{eq2:non-integer}, we have
\begin{align*}
G_\rho(\bx|Q_{\hX}^n)\doteq\exp\left\{\rho nR(P_{\bx},Q_{\hX}|\Delta)\right\}
\end{align*}
and thus,
\begin{align*}
  \lefteqn{\sum_{\bx\in\cX^n}P_{X^n}(\bx)G_\rho(\bx|Q_{\hX}^n)}\\
  &\doteq\sum_{Q_X\in\cP_n}\exp\left\{n\left[\rho R(Q_X,Q_{\hX}|\Delta)-D(Q_X\Vert P_X)\right]\right\}\\
  &\doteq\exp\left\{n\max_{Q_X}\left[\rho R(Q_X,Q_{\hX}|\Delta)-D(Q_X\Vert P_X)\right]\right\}.
\end{align*}

\section*{Acknowledgment}
This work was supported in part by JSPS KAKENHI Grant Number 18K04141.




\end{document}